\pdfoutput=1
\documentclass[lettersize,journal]{IEEEtran}
\usepackage{amsmath,amsfonts}
\usepackage{algorithmic}

\usepackage{algorithm}
\usepackage{array}
\usepackage[caption=false,font=normalsize,labelfont=sf,textfont=sf]{subfig}
\usepackage{textcomp}
\usepackage{stfloats}
\usepackage{url}
\usepackage{verbatim}
\usepackage{multirow}
\usepackage{graphicx}
\usepackage[hidelinks]{hyperref}
\usepackage{booktabs}
\usepackage{graphicx}
\begin{document}

\bstctlcite{IEEEexample:BSTcontrol}
\title{A High-Frequency Focused Network for Lightweight Single Image Super-Resolution}

\author{Xiaotian Weng, Yi Chen, Zhichao Zheng, Yanhui Gu, Junsheng Zhou, and Yudong Zhang, ~\IEEEmembership{Senior Member,~IEEE}
\thanks{The work is supported in part by the Natural Science Foundation of China (Nos. 22033002); in part by MRC, UK (MC\_PC\_17171); in part by Royal Society, UK (RP202G0230); in part by BHF, UK (AA/18/3/34220); in part by Hope Foundation for Cancer Research, in part by UK (RM60G0680); in part by GCRF, UK (P202PF11); in part by Sino-UK Industrial Fund, UK (RP202G0289); in part by LIAS, UK (P202ED10, P202RE969); in part by Data Science Enhancement Fund, UK (P202RE237); in part by Fight for Sight, UK (24NN201); in part by Sino-UK Education Fund, UK (OP202006) and in part by BBSRC, UK (RM32G0178B8). (Corresponding author: Yi Chen and Yudong Zhang.)}
\thanks{Xiaotian Weng, Yi Chen, Zhichao Zheng, Yanhui Gu and Junsheng Zhou are with the College of Computer and Electronic Information \& College of Artificial Intelligence at Nanjing Normal University, Nanjing 210023, China (e-mail: 212202045@njni.edu.cn; cs\_chenyi@njnu.edu.cn; zheng\_zhichaoX@163.com; gu@njnu.edu.cn; zhoujs@njnu.edu.cn)}
\thanks{Yudong Zhang is with the School of Computing and Mathematical Sciences, University of Leicester, LE17 RH Leicester, U.K. (e-mail: yudongzhang@ieee.org).}}

\markboth{Journal of \LaTeX\ Class Files,~Vol.~14, No.~8, August~2021}%
{Shell \MakeLowercase{\textit{et al.}}: A Sample Article Using IEEEtran.cls for IEEE Journals}

\IEEEpubid{0000--0000/00\$00.00~\copyright~2021 IEEE}

\maketitle

\begin{abstract}
  Lightweight neural networks for single-image super-resolution (SISR) tasks have made substantial breakthroughs in recent years. Compared to low-frequency information, high-frequency detail is much more difficult to reconstruct. Most SISR models allocate equal computational resources for low-frequency and high-frequency information, which leads to redundant processing of simple low-frequency information and inadequate recovery of more challenging high-frequency information. We propose a novel High-Frequency Focused Network (HFFN) through High-Frequency Focused Blocks (HFFBs) that selectively enhance high-frequency information while minimizing redundant feature computation of low-frequency information. The HFFB effectively allocates more computational resources to the more challenging reconstruction of high-frequency information. Moreover, we propose a Local Feature Fusion Block (LFFB) effectively fuses features from multiple HFFBs in a local region, utilizing complementary information across layers to enhance feature representativeness and reduce artifacts in reconstructed images. We assess the efficacy of our proposed HFFN on five benchmark datasets and show that it significantly enhances the super-resolution performance of the network. Our experimental results demonstrate state-of-the-art performance in reconstructing high-frequency information while using a low number of parameters.
\end{abstract}

\section{Introduction}
\IEEEPARstart{S}{ingle-image} super-resolution (SISR) is a fundamental low-level task in computer vision that involves generating high-resolution (HR) images from low-resolution (LR) inputs \cite{taiSuperResolutionUsing2010,gaoContextPatchRepresentationLearning2022,DBLP:conf/cvpr/SunXS08}, which goal is to reconstruct missing details and enhance image quality using advanced techniques in image processing and machine learning. This is a challenging assignment because high-frequency information is often lost during the downsampling process. Additionally, the ambiguity in mapping the LR image to its corresponding HR counterpart makes the problem ill-posed \cite{DBLP:conf/cvpr/TaiLBL10,DBLP:conf/cvpr/HeS11,DBLP:conf/cvpr/KimLL16a}. SISR has widespread applications in medical imaging \cite{shiCardiacImageSuperResolution2013} and video surveillance \cite{rastiConvolutionalNeuralNetwork2016}. In recent years, many SISR approaches have been proposed to attack this issue.

With the rise of deep neural networks, many SISR methods based on deep learning have emerged with impressive achievements, revolutionizing the field of SISR. Dong et al. made significant contributions to image super-resolution with their pioneering SRCNN \cite{dongLearningDeepConvolutional2014g}, which employed only three convolutional neural networks (CNN) layers yet achieved far better performance than traditional methods. Building upon residual learning \cite{heDeepResidualLearning2016}, Kim et al. proposed VDSR \cite{kimAccurateImageSuperResolution2016} to increase network depth to 20 layers further and utilize residual blocks to accelerate model convergence and improve performance. The groundbreaking EDSR \cite{limEnhancedDeepResidual2017} stacked the number of network layers to over 60, and its remarkable performance fully demonstrated the profound impact of network depth on image reconstruction quality. With the influx of attention mechanisms \cite{huSqueezeandExcitationNetworks2018,zhangImageSuperResolutionUsing2018,DBLP:conf/cvpr/DaiCZXZ19,DBLP:conf/eccv/NiuWRZYWZCS20,DBLP:conf/cvpr/MeiFZ21} and Transformer models \cite{dosovitskiyImageWorth16x162021,DBLP:journals/cvm/WangXLFSLLLS22}, the behavior of the SISR model has further enhanced.Zhang et al. introduced channel attention to adjust the weight of the model amongst different channels, by designing a residual structure to achieve RCAN \cite{zhangImageSuperResolutionUsing2018}. On the other hand, Liang et al. introduced swinTransformer into the field of image restoration and proposed swinIR \cite{liangSwinIRImageRestoration2021a}, which realized state-of-the-art performance in the SISR domain with Transformer's powerful expressive capability. Recently, the performance of HAT \cite{chenActivatingMorePixels2022a} proposed by Chen et al. is even more impactful. However, the impressive expressive power of these models often comes at a high computational cost, regardless of whether they use an attention mechanism or Transformer.
\IEEEpubidadjcol

As network performance increases, the number of model parameters and computational effort also escalate significantly, rendering the practical application of super-resolution networks infeasible. For SISR on mobile devices, a balance between performance and computational cost must be struck. As a result, both academia and industry \cite{DBLP:conf/eccv/ZhangDLTLTWZHXL20} are gradually shifting their focus to developing lightweight super-resolution methods that can achieve comparable results with much fewer parameters and lower computational effort. DRCN \cite{kimDeeplyRecursiveConvolutionalNetwork2016} and DRRN \cite{taiImageSuperResolutionDeep2017} achieve super-resolution through recursive network structure, and greatly reduce the number of parameters through a parameter-sharing strategy. However, the recursive learning approach brings an increase in network depth and computation while reducing the number of parameters, which is actually not as efficient as the SISR model. Therefore, the lightweight SISR model requires a lightweight and efficient structure. Ahn et al. proposed CARN \cite{ahnFastAccurateLightweight2018} to achieve fast, accurate, and efficient recovery of HR images by cascading local and global network layer information. Hui et al. proposed IMDN \cite{huiLightweightImageSuperResolution2019} to enhance the capability and efficiency of SISR by introducing information multi-distillation blocks for feature extraction and refinement. Liu et al. further improved the IMDN by introducing a noval feature distillation connections (FDCs) and proposed RFDN \cite{liuResidualFeatureDistillation2020}.  Li et al. designed BSRN \cite{liBlueprintSeparableResidual2022a} by introducing blueprint separable convolution to improve RFDN again, which improves the performance with reduced model size. Lu et al. proposed ESRT \cite{luTransformerSingleImage2022} to achieve lightweight SISR models with the Transformer. They designed an efficient Transformer structure that is used in a hybrid network, combining both CNN and Transformer structures. This hybrid approach further improves lightweight SISR models while maintaining their efficiency and reducing computational costs. The ESRT achieves the best performance by effectively leveraging the strengths of both CNN and Transformer structures in image reconstruction.
\begin{figure}[!t]
  \centering
  \includegraphics[width=3.2in]{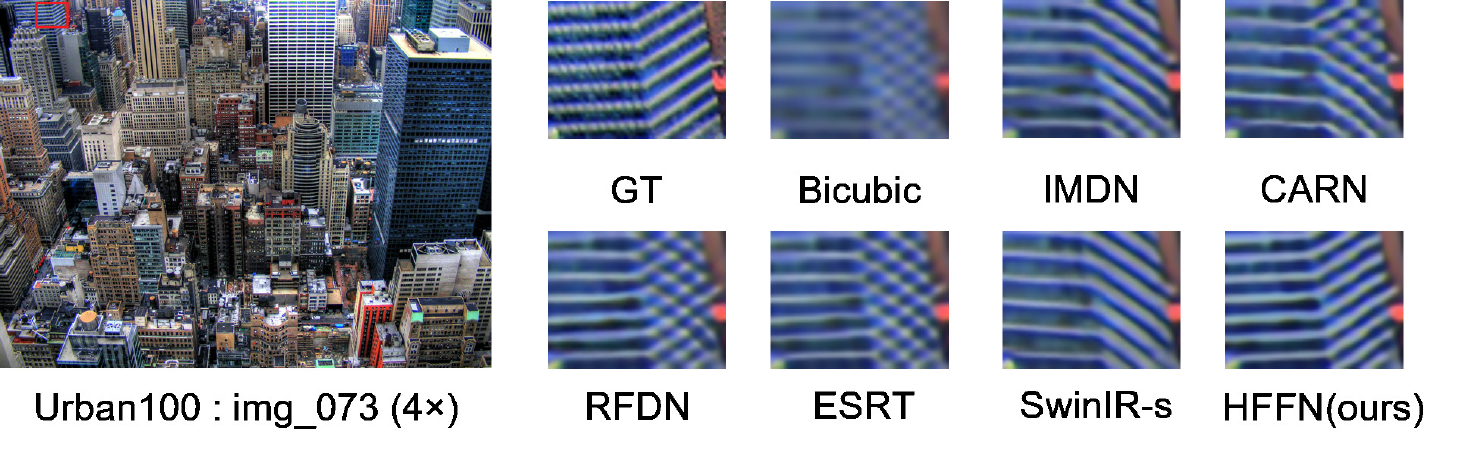}
  \caption{Visual comparison for images generated by state-of-the-art lightweight models. Our HFFN is able to generate the most correct images in regions with more high high-frequency features.}
  \label{fig_1}
  \end{figure}
It has been demonstrated in large SISR models \cite{zhangImageSuperResolutionUsing2018,chenActivatingMorePixels2022a} that high-frequency information in images is significant for super-resolution tasks, and insufficient recovery of high-frequency details usually leads to blurred images with poor sharpness. Low-frequency features do not suffer a significant loss during image degradation, so low-frequency information can be obtained directly from the input \cite{behjatiFrequencyBasedEnhancementNetwork2022a}. Despite the recent great progress in lightweight SISR models, since most lightweight SISR models handle high- and low-frequency features as equal, which means that both enjoy the same computational resources, however, high-frequency features require higher computational resources compared to low-frequency information such as image texture and color. Allocating a large of computational resources to low-frequency information is a redundant computation \cite{wangLightweightFeatureDeredundancy2023}. Therefore, it can be seen from Fig. \ref{fig_1} that the existing lightweight model still has difficulty in correctly recovering the high-frequency details in the image. In this paper, we propose High-Frequency Focused Block (HFFB), which has two branches, where the high-frequency enhancement branch uses a filter structure to extract and enhance high-frequency features, and the low-frequency de-redundant branch uses a feature de-redundancy structure to minimize the redundant computation of low-frequency features.  Furthermore, we design a noval Local Feature Fusion Block (LFFB) to improve further all features extracted by HFFBs. By utilizing these novel blocks, our proposed High-Frequency Focused Network (HFFN) achieves unprecedented accuracy and efficiency in SISR, effectively recovering high-frequency details while significantly reducing computational costs. The main contributions of this paper are as follows:

\begin{enumerate}
\item{We designed a noval high-frequency focused network (HFFN) to better reconstruct images with sharp edges or textures by allocating more resources to challenging high-frequency information.}
\item{We propose a high-frequency focused block (HFFB) to enhance the high-frequency information while minimizing the redundant feature computation of low-frequency information. In addition, we design a local feature fusion block (LFFB) to enhance the feature information extracted from the HFFB.}
\item{Our model has been extensively evaluated through various experiments, and the results consistently demonstrate its superiority in terms of recovering high-frequency details.}
\end{enumerate}

\section{Related Works}
\subsection{Lightweight Super-Resolution Model}
Since SRCNN \cite{dongLearningDeepConvolutional2014g} first used convolutional neural convolution for super-resolution models, more and more deep learning-based methods have been proposed one after another and achieved promising results. At the same time, the size and the computational effort of the models are gradually expanding, which is unfavorable to the application of super-resolution in real life. Recently, there has been a growing interest in lightweight super-resolution networks, and researchers are proposing more and more models that are both accurate and efficient. Kim et al. and Tai et al. devised the DRCN \cite{kimDeeplyRecursiveConvolutionalNetwork2016} and DRRN \cite{taiImageSuperResolutionDeep2017} models, respectively, utilizing a recursive structure and parameter-sharing mechanism to drastically reduce the number of model parameters. However, the recursive structure did not reduce the computational effort. The CARN \cite{ahnFastAccurateLightweight2018} proposed by Ahn et al. improved the performance and efficiency of super-resolution using cascaded networks. Then Hui et al. implemented a lightweight information multi-distillation network (IMDN) \cite{huiLightweightImageSuperResolution2019} by information distillation, which can utilize more useful features. With the popularity of attentional mechanisms, a variety of lightweight models \cite{9427111,8708220,9072448,zhaoEfficientImageSuperResolution2020a} based on them have likewise arisen. Zhao et al. \cite{zhaoEfficientImageSuperResolution2020a} proposed a novel approach for super-resolution by introducing pixel attention (PA) and designing a network architecture with a self-calibrating mechanism \cite{liuImprovingConvolutionalNetworks2020}, which achieves comparable performance with a much smaller number of parameters. The ECBSR \cite{DBLP:conf/mm/ZhangZZ21} was cleverly designed by zhang et al. using a re-parameterization technique to achieve super-resolution for edge devices. Inspired by swinTransformer \cite{liuSwinTransformerHierarchical2021}, Liang et al. \cite{liangSwinIRImageRestoration2021a} proposed a lightweight architecture, swinIR-s, which exploits the powerful feature representation capability of Lu et al.\cite{luTransformerSingleImage2022} designed ESRT, which effectively combines the advantages of CNN and Transformer and further improves the performance of lightweight SISR networks. Although the recent lightweight SISR models have achieved good results, they still have some problems in recovering high-frequency details of images because they do not focus too much on the high-frequency information of images.
\subsection{Frequency-Based Super-Resolution Model}
In the field of SISR, high-frequency information such as edge plays a critical role in the image restoration process. Li et al. designed a super-resolution feedback network (SRFBN) \cite{liFeedbackNetworkImage2019} that progressively refines high-frequency features by continuously transforming the image between high- and low-resolution images. Li et al. \cite{liMultiscaleResidualNetwork2018} separated high-frequency information from low-frequency information by introducing a new octave convolution (OC) and achieved information update and frequency communication between high- and low-frequency information during the convolution process. FilterNet \cite{8672189} used a gating unit to suppress low-frequency information to better learn more useful information. Fang et al. \cite{fangSoftEdgeAssistedNetwork2020} designed SeaNet, a two-stage super-resolution network that reconstructs the image's soft edges in the first stage, then combines high- and low-frequency features in the second stage. By introducing explicit high-frequency a priori information,   
\begin{figure*}[!t]
\centering
  \subfloat[]{\includegraphics[width=7.0in]{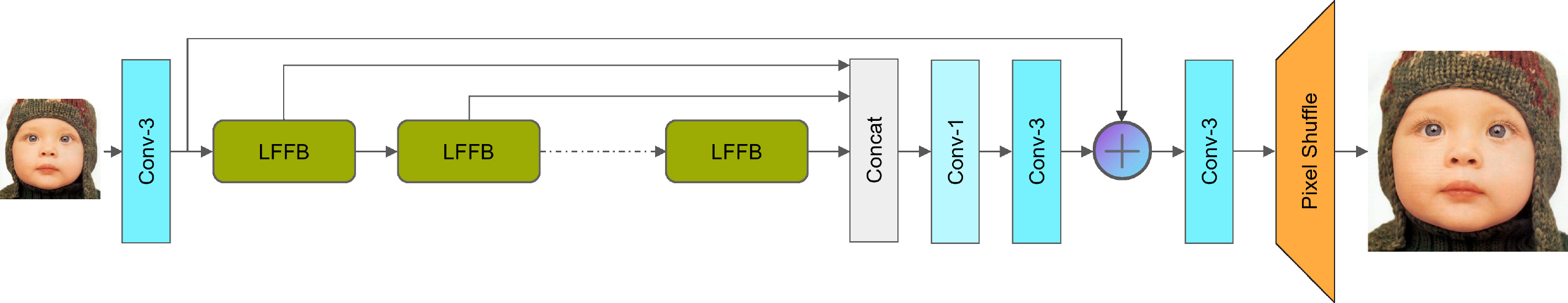}%
  \label{fig_first_case}}\\
  \subfloat[]{\includegraphics[width=4.0in]{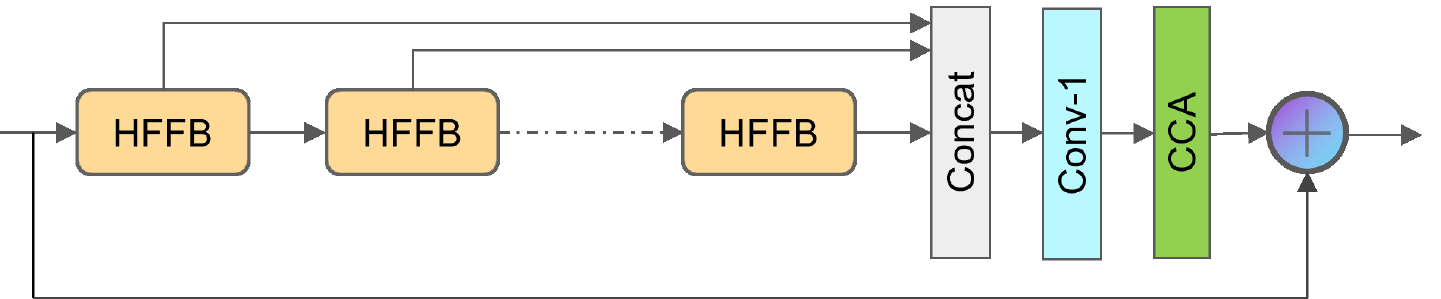}%
  \label{fig_second_case}}\hfill
  \subfloat[]{\includegraphics[width=1.5in]{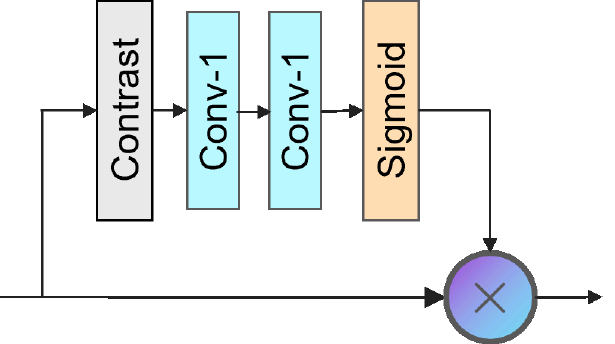}%
  \label{fig_thrid_case}}\hfill
  \subfloat[]{\includegraphics[width=1.5in]{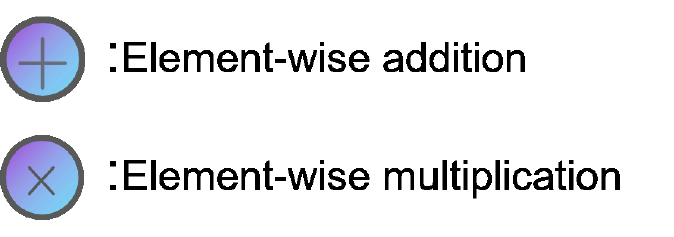}%
  \label{fig_four_case}}
  \caption{The framework of HFFN:(a) The overall network architecture of HFFN.(b) Local feature fusion block (LFFB).(c) Contrast channel attention (CCA).(d) Figure legend.}
  \label{fig_2}
  \end{figure*}SeaNet is able to reconstruct the sharp high-frequency details of an image remarkably well. Yang et al. \cite{yangDeeplyRecursiveLow2020} proposed DRFFN, a deep recursive low- and high-frequency fusion network that utilizes a parallel branching structure to extract both low- and high-frequency feature from image. Behjati et al. \cite{behjatiFrequencyBasedEnhancementNetwork2022a} proposed a novel frequency-based enhancement network (FENet), which enhances the high-frequency information of an image by a feature enhancement block. Lu et al. recently proposed ESRT \cite{luTransformerSingleImage2022}, a hybrid CNN-Transformer network that introduces a High-frequency Filtering Module (HFM) in its CNN part. The HFM is a high-pass filter-like structure that can effectively capture the high-frequency information of images. While the aforementioned methods have improved the network's performance by paying attention to high-frequency details, most do so at the cost of increased computational complexity. In this paper, we propose a high-frequency focused block that reallocates computational resources from low-frequency information to high-frequency information, enabling the reconstruction of better image details without increasing the overall complexity of the model.

\section{Methodology}
\subsection{Network Architecture}
The network framework is illustrated in Fig. \ref{fig_2}. Our HFFN contains four main parts: shallow feature extraction (SFE), high-frequency feature extraction consisting of LFFB, feature fusion, and finally the image reconstruction part. In this work, we first define the input of the HFFN as ${I_{LR}}$ and the output as ${I_{SR}}$ . then, the first part of the network extracts the shallow features of the input ${I_{LR}}$ using a convolutional layer with a convolutional kernel size of $3 \times 3$ . The shallow features can be expressed as:

\begin{equation}
\label{deqn_ex1a}
{F_0} = {H_{SFE}}({I_{LR}})
\end{equation}
where ${H_{SFE}}( \cdot )$ refers to the shallow feature extraction layer and ${F_0}$ indicates that the shallow features extracted from the ${I_{LR}}$ are then used as input for deep feature extraction consisting of multiple LFFBs.
\begin{equation}
\label{deqn_ex2a}
{F_{\rm{n}}} = H_{LFFB}^n(H_{LFFB}^{n - 1}(...(H_{{\rm{LFFB}}}^1({{\rm{F}}_0})...))
\end{equation}
where $H_{LFFB}^n( \cdot )$ represents the mapping of the $n{\rm{ - th}}$ LFFB module and ${F_{\rm{n}}}$ is the output of the $n{\rm{ - th}}$ LFFB module. In this paper, we set n to 6 based on extensive experiments. To better utilize the feature information extracted from each LFFB, we immediately cascade the outputs of all the LFFBs together and fuse the features by a $1 \times 1$ convolution, and then refine the features by a $3 \times 3$ convolution. The feature fusion equation is:
\begin{equation}
\label{deqn_ex3a}
{F_{fused}} = Con{v_{3 \times 3}}(Con{v_{1 \times 1}}(Concat({F_1},,,{F_n})))
\end{equation}
where $Concat( \cdot )$ refers to the cascade in the channel dimension, $Con{v_{1 \times 1}}( \cdot )$ means the convolution operation with a convolution kernel of $1 \times 1$, and $Con{v_{3 \times 3}}( \cdot )$ denotes the convolution operation with a convolution kernel of $3 \times 3$. ${F_{fused}}$ represents the fused features.

Next, our extracted shallow features   are input to the reconstruction layer along with the fused features. The reconstruction layer consists of a convolutional kernel of size $3 \times 3$ and a Pixel-Shuffle layer to scale the LR to HR image size\cite{shiRealTimeSingleImage2016}. As a result, we obtain:
\begin{equation}
  \label{deqn_ex4a}
  {I_{SR}} = {H_{HFFN}}({I_{LR}}) = {H_{ps}}(Con{v_{3 \times 3}}({F_0} + {F_{fused}}))
  \end{equation}
where ${H_{HFFN}}( \cdot )$ denotes our entire network mapping, ${H_{ps}}( \cdot )$ represents the pixel-shuffle layer, and $Con{v_{3 \times 3}}( \cdot )$ denotes the convolutional kernel size of $3 \times 3$ convolutional operations.

For training, we are given a set of images, i.e., $\{ I_{LR}^i,I_{HR}^i\} _{i = 1}^N$, where $N$ means the number of training set pairs of images. The most common ${L_1}$ loss is used to optimize our model. The loss function can be expressed as:
\begin{equation}
  \label{deqn_ex5a}
  {L_1}(\theta ) = \frac{1}{N}\sum\limits_{i = 1}^N {{{\left\| {{H_{HFFN}}(I_{LR}^i) - I_{HR}^i} \right\|}_1}}
  \end{equation}
  where $\theta$ means the parameters that can be learned during the training process and ${\left\|  \cdot  \right\|_1}$ denotes the ${L_1}$ norm.
\begin{figure*}[!t]
  \centering
  \subfloat{\includegraphics[width=7.0in]{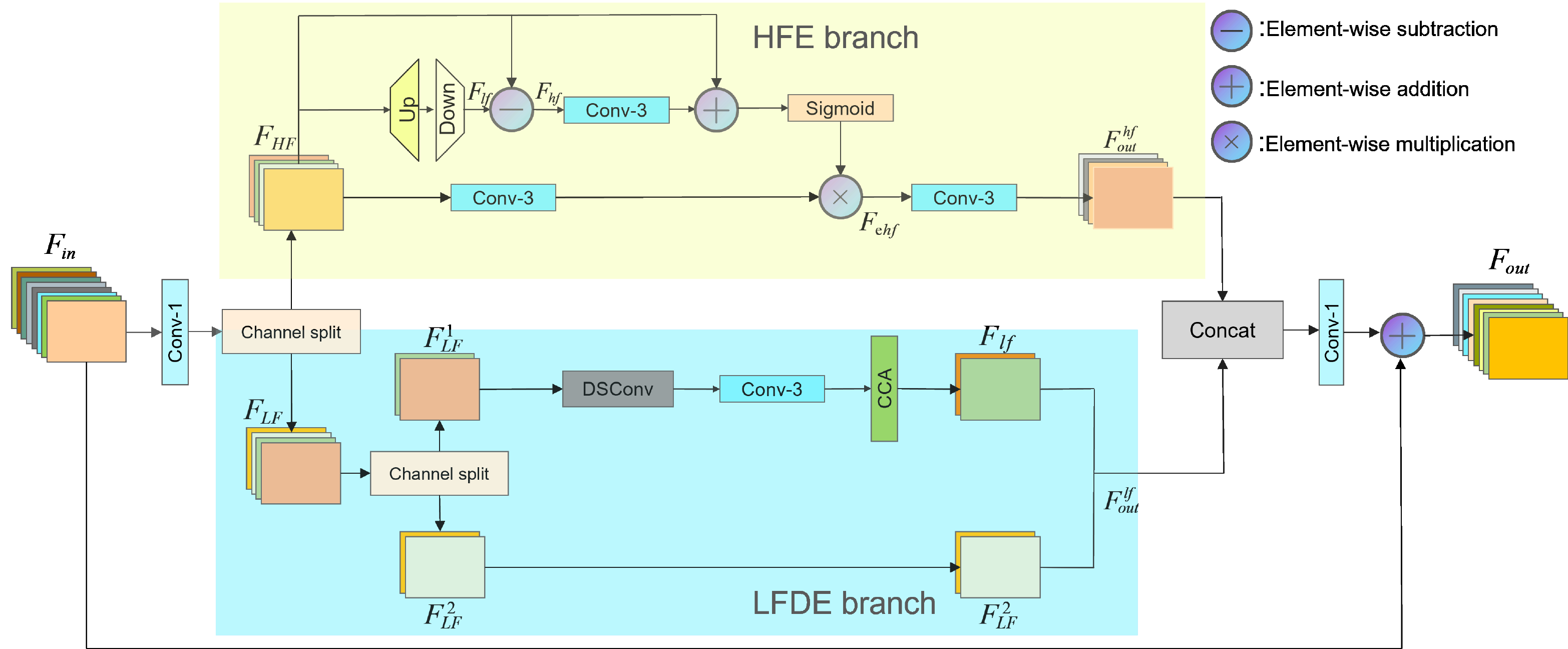}%
  \label{fig_2_case}}
  \caption{The illustration of the high-frequency focus block (HFFB) we designed, where the yellow part is the high-frequency enhancement branch and the blue part is the low-frequency de-redundancy branch.}
  \label{fig3}
  \end{figure*}
\subsection{High-Frequency Focused Block (HFFB)}
An image includes low-frequency features like color and texture and sharp high-frequency features like edges. We usually think that the majority of information lost during image degradation is high-frequency features, and low-frequency features is retained in the degraded LR image. Therefore, we design a high-frequency focus block (HFFB), which can focus on reconstructing high-frequency information and minimize the redundant computation of low-frequency information. 

As demonstrated in Fig. \ref{fig3}, the initial feature information first undergoes a $1 \times 1$ convolution for initial feature extraction, and the extracted information flows from the channel split into two branches, the upper branch is designed to extract and reinforce the high-frequency information, and the lower branch is used to reduce the redundant feature processing of the low-frequency information. Thus, we can obtain:

\begin{equation}
  \label{deqn_ex6a}
  {F_{HF}},{F_{LF}} = {f_{split}}(Con{v_{1 \times 1}}({F_{in}}))
  \end{equation}
where $Con{v_{1 \times 1}}( \cdot )$ denotes the $1 \times 1$ convolutional operation, ${f_{split}}( \cdot )$ denotes the channel segmentation processing, ${F_{HF}}$ and ${F_{LF}}$ denotes the inputs of the next two branches, and ${F_{in}}$ represents the input of HFFB.

Next, the first pathway is used to extract and reinforce high-frequency information,We call it high-frequency enhancement branch(HFE). Inspired by high pass filters, to extract high-frequency information, the input features are first upsampled by a deconvolution, and then a downsampling layer is used to downsample the image to achieve a class blurring effect, at which time the extracted features are low-frequency information. We subtract the extracted low-frequency information from the input features one by one to gain the high-frequency information:
\begin{equation}
  \label{deqn_ex7a}
  {F_{lf}} = {H_{ds}}({H_{us}}({F_{HF}}))
  \end{equation} 
\begin{equation}
  \label{deqn_ex7b}
  {F_{hf}} ={F_{HF}}  -  {F_{lf}} 
  \end{equation} 
  where ${H_{us}}( \cdot )$ denotes the upsampling layer, ${H_{ds}}( \cdot )$ represents the downsampling layer, ${F_{lf}}$ denotes the low-frequency information obtained downsampling, and ${F_{hf}}$ denotes the high-frequency information through element-by-element subtraction, respectively. Further, we perform high-frequency information feature enhancement:
  \begin{equation}
    \label{deqn_ex8a}
    {F_{{\rm{e}}hf}} = \sigma (Con{v_{3 \times 3}}({{\rm{F}}_{hf}}) + {F_{HF}}) \cdot Con{v_{3 \times 3}}({F_{HF}}))
    \end{equation} 
\begin{equation}
    \label{deqn_ex8va}
    F_{out}^{hf} = Con{v_{3 \times 3}}({F_{{\rm{e}}hf}})
    \end{equation} 
where $Con{v_{3 \times 3}}( \cdot )$ represents the convolution operation with a convolution size of $3 \times 3$, $\sigma (\cdot )$ denotes the sigmoid function, ${F_{{\rm{e}}hf}}$ is the feature information after enhancing the high-frequency, ${F_{out}^{hf}}$ is the output of HFE.

The second branch is responsible for reducing the redundant computation of low-frequency features, We call it low-frequency de-redundant branch(LFDE). Inspired by CSPNET \cite{wangCSPNetNewBackbone2020}, since the image's low-frequency features can be basically obtained from LR, in order to minimize the computational effort, we perform another channel segmentation on the input information. Half of the channel information is transferred directly to the output, while the other half is subjected to some efficient processing. In this we use a depthwise separable convolution(DSConv) \cite{cholletXceptionDeepLearning2017}, a normal convolution with a convolution kernel size of $3 \times 3$, and finally a contrast channel attention (CCA). Depth-separable convolution consists of $3 \times 3$ channel-wise convolution and $1 \times 1$ point-wise convolution, which can reduce the parameter count and computational effort of the model. The CCA consists of a Contrast layer and two $1 \times 1$ convolutional layers. This structure can be expressed as:
\begin{equation}
  \label{deqn_ex9a}
  F_{LF}^1,F_{LF}^2 = {f_{split}}({F_{LF}})
  \end{equation}  
\begin{equation}
\label{deqn_ex10a}
{F_{lf}} = {H_{CCA}}(Con{v_{3 \times 3}}({H_{PW}}({H_{DW}}(F_{LF}^1))))
\end{equation} 
\begin{equation}
\label{deqn_ex10b}
F_{out}^{lf} = Concat({F_{lf}},F_{LF}^2)
\end{equation} 
where ${H_{DW}}( \cdot )$ represents the channel convolution operation, ${H_{PW}}( \cdot )$ denotes the point convolution operation, and ${H_{CCA}}( \cdot )$ denotes the contrast channel attention. $F_{LF}^1$, $F_{LF}^2$ represents the input flowing to the two branches after channel segmentation, and ${F_{lf}}$  is the extracted low-frequency information, $F_{out}^{lf}$ is the output of the low frequency de-redundant branch.

Then, we fuse the output features of the two branches by convolutional layers, and finally, use the residual join to obtain the final output feature map:
\begin{equation}
\label{deqn_ex11a}
F_{out} = Con{v_{1 \times 1}}(Concat({F_{out}^{hf}},F_{out}^{lf})) + {F_{in}}
\end{equation}   
where $Concat( \cdot )$ represents the cascade operation, $Con{v_{1 \times 1}}( \cdot )$ is the convolution operation with a convolution kernel of $1 \times 1$, and ${F_{out}}$ represents the output feature information of HFFB.

With our HFFB, we can effectively devote more computational resources to more difficult to reconstruct high-frequency features. With this feature of HFFB, our network can recover finer images while maintaining a light weight. 
\subsection{Local Feature Fusion Block (LFFB)}
Since feature information is lost during transmission in the network, we designed a noval block in order to better utilize all feature information extracted from each HFFB. We called it LFFB. It does the fusion as well as refinement of the information extracted from the HFFB of a local area and then transmits the fused information to the next LFFB and the final feature fusion stage. Remarkably, we used contrast channel attention to adjust the feature information after fusion before the residuals were connected. As shown in Fig. \ref{fig_2}, the LFFB mainly consists of m HFFBs, a connection layer, a convolutional layer with a convolutional kernel, and a CCA layer.
\begin{table}[!t]
 \caption{Average PSNR obtained when either low-frequency de-redundant branch or high-frequency enhancement branch is deactivated inside the HFFB on five benchmark datasets with scale factor ×4. The value of the difference from the baseline is marked in parentheses.
\label{table1}}
 \centering
 \begin{tabular}{p{4.19em}ccc}
    \toprule
    Method & w/o LFDE & w/o HFE & HFFN \\
    \midrule
    Params & 880K  & 803K  & 867K \\
    \midrule
    Set5  & 32.21 (-0.17dB) & 32.28 (-0.10dB) & 32.38 \\
    Set14 & 28.60 (-0.10dB) & 28.56 (-0.14dB) & 28.70 \\
    B100  & 27.60 (-0.06dB) & 27.57 (-0.09dB) & 27.66 \\
    Urban100 & 26.11 (-0.30dB) & 26.12 (-0.29dB) & 26.41 \\
    Manga109 & 30.53 (-0.28dB) & 30.48 (-0.33dB) & 30.81 \\
    \bottomrule
  \end{tabular}
  \end{table}
\begin{equation}
\label{deqn_ex12a}
F_{lffb}^0 = F_{lffb}^{in}
\end{equation}  
\begin{equation}
\label{deqn_ex13a}
F_{lffb}^k = {H_{HFFB}}(F_{lffb}^{k - 1}),k = 1,2,...,m
\end{equation}  
\begin{equation}
\label{deqn_ex14a}
{F_{concat}} = Concat(F_{lffb}^1,F_{lffb}^1,...,F_{lffb}^m)
\end{equation} 
\begin{equation}
  \label{deqn_ex15a}
  F_{lffb}^{out} = {H_{CCA}}(Con{v_{1 \times 1}}({F_{concat}})) + F_{lffb}^{in}
  \end{equation} 
where $F_{lffb}^k$ represents the output of the $k{\rm{ - th}}$ HFFB in the LFFB, $F_{concat}$ indicates the feature after the channel join operation, $F_{lffb}^{in}$ and $F_{lffb}^{out}$ represent the input and output of the LFFB. $Con{v_{1 \times 1}}( \cdot )$ represents a $1 \times 1$ convolution operation, $Concat( \cdot )$ represents the channel join operation, ${H_{HFFB}}( \cdot )$ represents the feature mapping of the HFFB, and ${H_{{\rm{CCA}}}}( \cdot )$ denotes the contrast channel attention.
\section{Experiments}
\subsection{Datasets and Metrics}

We used the most widely used DIV2K \cite{timofteNTIRE2018Challenge2018} for training, which has 900 images, of which 800 are used as the training set, and 100 are the validation set. LR images were aquired by double triple down-sampling We also tested our model on five common benchmark datasets: Set5 \cite{bevilacquaLowComplexitySingleImageSuperResolution2012}, Set14 \cite{yangImageSuperResolutionSparse2010}, BSD100 \cite{martinDatabaseHumanSegmented2001}, Urban100 \cite{huangSingleImageSuperresolution2015}, and Manga109 \cite{aizawaBuildingMangaDataset2020}. We evaluated our reconstructed images using PSNR and SSIM \cite{matsuiSketchbasedMangaRetrieval2017} as metrics, where both metrics are of higher values indicating better quality of the recovered images.
\subsection{Experimental Details}
For training our model, we input 48 LR images of size $48X48$ per batch into the model and use flips and rotations for data enhancement. Meanwhile, we use Adam optimizer \cite{DBLP:journals/corr/KingmaB14} for training, with parameters set to 0.9 and 0.999, initial learning rate set to $6 \times {10^{ - 4}}$. The number of channels in our HFFN is set to 48. All models are implemented using Pytorch \cite{DBLP:conf/nips/PaszkeGMLBCKLGA19} and trained on 2 NVIDIA RTX 3080.
\subsection{Ablation Study}
\subsubsection{Study of high-frequency focused block (HFFB)}
HFFB is the basic component of our proposed network, which can allocate computational resources to the high-frequency features that need to be computed the most. To verify the validity of HFFB, we explore the effectiveness of each path of HFFB in Table \ref{table1}. Specifically, we run the following experiments: (1): activating only the high-frequency enhancement branch (2) activating only the low-frequency de-redundant branch, where deactivation uses $3 \times 3$ convolutional layers instead of paths. Meanwhile, to ensure fairness, we choose to add HFFB blocks to control the three model parameters at about 800K, while training the three models and our HFFN for 500 epochs.

\begin{figure}[t]
    \centering
    \includegraphics[width=3.2in]{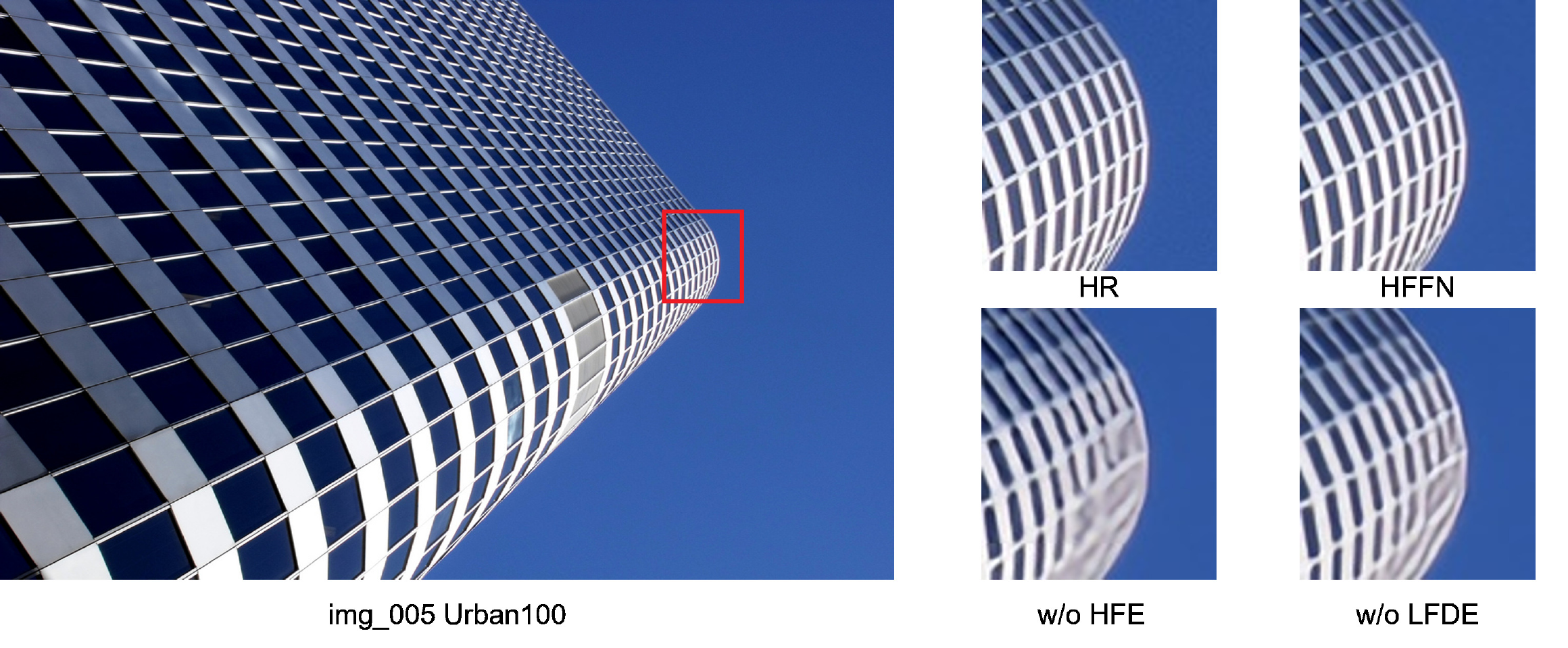}
    \caption{Visual comparison chart of SR images obtained by disabling HFFN with different paths.}
    \label{fig_4}
    \end{figure}

 \begin{table}[t]
\caption{The average PSNR obtained when the DSConv of the low-frequency de-redundant branch of the HFFB is deactivated with the CCA module on the five benchmark data sets with scale factor × 4. The value of the difference from the baseline is marked in parentheses.\label{table2}}
\centering
\begin{tabular}{p{4.19em}ccc}
  \toprule
  Method & w/o DSConv & w/o CCA & HFFN \\
  \midrule
  Params & 898K  & 862K  & 867K \\
  \midrule
  Set5  & 32.38 (-0.00dB) & 32.34 (-0.04dB) & 32.38 \\
  Set14 & 28.68 (-0.02dB) & 28.71 (+0.01dB) & 28.70 \\
  B100  & 27.65 (-0.01dB) & 27.63 (-0.03dB) & 27.66 \\
  Urban100 & 26.35 (-0.06dB) & 26.27 (-0.14dB) & 26.41 \\
  Manga109 & 30.80 (-0.01dB) & 30.74 (-0.07dB) & 30.81 \\
  \bottomrule
\end{tabular}
\end{table} 
  From Table \ref{table1}, it can be observed that a large degradation in model performance occurs when both thelow-frequency de-redundant branch is deactivated, and the high-frequency enhancement branch is deactivated. Due to the fact that when the low-frequency de-redundant branch is deactivated, the model focuses too much on the high-frequency details and loses the low-frequency information, resulting in the insufficient recovery of the overall contour of the image. When disabling the high-frequency enhancement branch, the model does not enhance the high-frequency information by processing both high and low-frequency information, resulting in a picture with insufficient detail and too much blur. To better confirm this, we provide a visual comparison of the above models in Fig. \ref{fig_4}.
\subsubsection{Effectiveness of high-frequency enhancement branch}
\begin{figure}[t]
  \centering
  \includegraphics[width=2.5in]{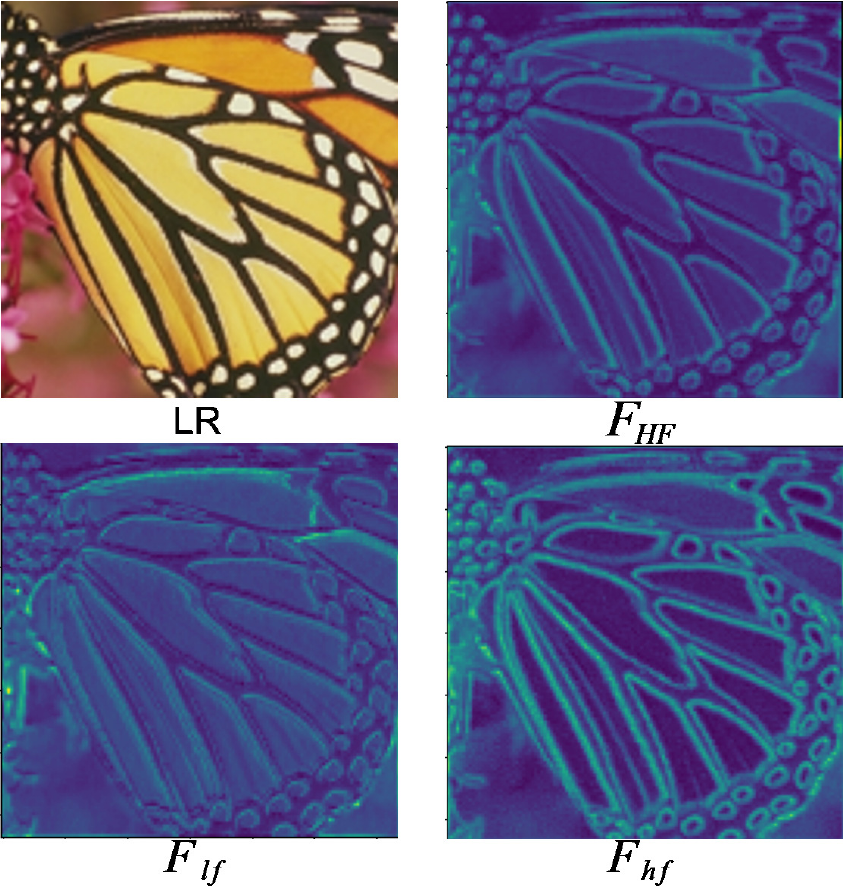}
  \caption{Visualization of the feature activation map for the original input information, low-frequency information, and high-frequency information, which are generated by the HFE of the HFFB.}
  \label{fig_5}
  \end{figure}

To confirm the effectiveness of our high-frequency enhancement branch, the visualized activation feature maps of the original feature(${F_{HF}}$), the blurred low-frequency feature(${F_{lf}}$), and the extracted high-frequency feature(${F_{hf}}$) is provided in Fig. \ref{fig_5}. Where LR is the input picture of HFFN. As can be observed from the images, our HFFN extracts the vast majority of the high-frequency information.
\begin{table}[t]
  \caption{The average PSNR obtained when the DSConv of the low-frequency de-redundant branch of the HFFB is deactivated with the CCA module on the five benchmark data sets with scale factor × 4. The value of the difference from the baseline is marked in parentheses.\label{table3}}
  \centering
  \begin{tabular}{p{4.19em}ccc}
    \toprule
    Number & 4 & 5 & 6 \\
    \midrule
    Params & 705K  & 867K  & 1026K  \\
    \midrule
    Set5  & 32.36 / 0.8972 & \pmb{32.49} / \pmb{0.8988} & 32.39 / 0.8977 \\
    Set14 & 28.68 / 0.7837 & \pmb{28.83} / \pmb{0.7870} & 28.80 / 0.7864 \\
    B100  & 27.63 / 0.7384 & \pmb{27.74} / \pmb{0.7413} & 27.71 / 0.7408 \\
    Urban100 & 26.29 / 0.7930 & \pmb{26.62} / \pmb{0.8014} & 26.57 / 0.8000 \\
    Manga109 & 30.69 / 0.9109 & 31.09 / 0.9134 & \pmb{31.11} / \pmb{0.9154} \\
    \bottomrule
  \end{tabular}
  \end{table}
\begin{table*}[!t]
  \centering
  \caption{A quantitative comparison of the average PSNR/SSIM with other advanced SISR models and the number of parameters. The best and second-best results are bolded and underlined, respectively. ’+’ means that the model used a self-ensemble strategy.\label{table6}}
    \begin{tabular}{p{7.625em}ccccccc}
    \toprule
    \multirow{2}[4]{*}{Method} & \multicolumn{1}{c}{\multirow{2}[4]{*}{Scale}} & \multirow{2}[4]{*}{Params} & Set5  & Set14 & BSDS100 & Urban100 & Manga109 \\
\cmidrule{4-8}    \multicolumn{1}{c}{} &       & \multicolumn{1}{c}{} & PSNR/SSIM & PSNR/SSIM & PSNR/SSIM & PSNR/SSIM & PSNR/SSIM \\
    \midrule
    VDSR \cite{kimAccurateImageSuperResolution2016}  & \multicolumn{1}{c}{\multirow{11}[2]{*}{×2}} & 666K  & 37.53 / 0.9587 & 33.03 / 0.9124 & 31.90 / 0.8960 & 30.76 / 0.9140 & 37.22 / 0.9750 \\
    CARN \cite{ahnFastAccurateLightweight2018}  &       & 1592K & 37.76 / 0.9590 & 33.52 / 0.9166 & 32.09 / 0.8978 & 31.92 / 0.9256 & 38.36 / 0.9765 \\
    IMDN \cite{huiLightweightImageSuperResolution2019}  &       & 694K  & 38.00 / 0.9605 & 33.63 / 0.9177 & 32.19 / 0.8996 & 32.17 / 0.9283 & 38.88 / 0.9774 \\
    RFDN \cite{liuResidualFeatureDistillation2020}  &       & 534K  & 38.05 / 0.9606 & 33.68 / 0.9184 & 32.16 / 0.8994 & 32.12 / 0.9278 & 38.88 / 0.9773 \\
    LatticeNet+ \cite{luoLatticeNetLightweightImage2020} &       & 756K  & 38.15 / 0.9610 & 33.78 / 0.9193 & 32.25 / 0.9004 & 32.29 / 0.9291 & - \\
    MAFFSRN \cite{muqeetMultiattentionBasedUltra2020} &       & 790K  & 38.07 / 0.9607 & 33.59 / 0.9177 & 32.23 / 0.9005 & 32.38 / 0.9308 & - \\
    SMSR \cite{wangExploringSparsityImage2021}  &       & 985K  & 38.00 / 0.9601 & 33.64 / 0.9179 & 32.17 / 0.8990 & 32.19 / 0.9284 & 38.76 / 0.9771 \\
    ESRT \cite{luTransformerSingleImage2022}  &       & 677K  & 38.03 / 0.9605 & 33.75 / 0.9184 & 32.25 / 0.9001 & 32.58 / 0.9318 & 39.12 / 0.9774 \\
    SwinIR-s \cite{liangSwinIRImageRestoration2021a} &       & 878K  & 38.14 / \underline{0.9611} & 33.86 / 0.9206 & 32.31 / 0.9012 & 32.76 / 0.9340 & 39.12 / 0.9783 \\
    HFFN (ours) &       & 851K  & \underline{38.17} / \underline{0.9611} & \underline{33.90} / \underline{0.9214} & \underline{32.32} / \underline{0.9014} & \underline{32.79} / \underline{0.9341} & \underline{39.33} / \underline{0.9784} \\
    HFFN+ (ours) &       & 851K  & \pmb{38.22} / \pmb{0.9613} & \pmb{34.02} / \pmb{0.9223} & \pmb{32.35} / \pmb{0.9017} & \pmb{32.97} / \pmb{0.9353}  & \pmb{39.46} / \pmb{0.9786} \\
    \midrule
    VDSR \cite{kimAccurateImageSuperResolution2016}  & \multicolumn{1}{c}{\multirow{11}[2]{*}{×3}} & 666K  & 37.53 / 0.9587 & 33.03 / 0.9124 & 31.90 / 0.8960 & 30.76 / 0.9140 & 37.22 / 0.9750 \\
    CARN \cite{ahnFastAccurateLightweight2018}  &       & 1592K & 34.29 / 0.9255 & 30.29 / 0.8407 & 29.06 / 0.8034 & 28.06 / 0.8493 & 33.43 / 0.9427 \\
    IMDN \cite{huiLightweightImageSuperResolution2019}  &       & 703K  & 34.36 / 0.9270 & 30.32 / 0.8417 & 29.09 / 0.8046 & 28.17 / 0.8519 & 33.61 / 0.9445 \\
    RFDN \cite{liuResidualFeatureDistillation2020}  &       & 541K  & 34.41 / 0.9273 & 30.34 / 0.8420 & 29.09 / 0.8050 & 28.21 / 0.8525 & 33.67 / 0.9449 \\
    LatticeNet+ \cite{luoLatticeNetLightweightImage2020} &       & 765K  & 34.53 / 0.9281 & 30.39 / 0.8424 & 29.15 / 0.8059 & 28.33 / 0.8538 & - \\
    MAFFSRN \cite{muqeetMultiattentionBasedUltra2020} &       & 807K  & 34.45 / 0.9277 & 30.40 / 0.8432 & 29.13 / 0.8061 & 28.26 / 0.8552 & - \\
    SMSR \cite{wangExploringSparsityImage2021}  &       & 993K  & 34.40 / 0.9270 & 30.33 / 0.8412 & 29.10 / 0.8050 & 28.25 / 0.8536 & 33.68 / 0.9445 \\
    ESRT \cite{luTransformerSingleImage2022}  &       & 770K  & 34.42 / 0.9268 & 30.43 / 0.8433 & 29.15 / 0.8063 & 28.46 / 0.8574 & 33.95 / 0.9455 \\
    SwinIR-s \cite{liangSwinIRImageRestoration2021a} &       & 886K  & 34.62 / 0.9289 & \underline{30.54} / \underline{0.8463} & 29.20 / \underline{0.8082} & \underline{28.66} / \underline{0.8624} & 33.98 / 0.9478 \\
    HFFN (ours) &       & 857K  & \underline{34.67} / \underline{0.9292} & \underline{30.54} / 0.8460 & \underline{29.22} / \underline{0.8082} & 28.64 / 0.8614 & \underline{34.22} / \underline{0.9480} \\
    HFFN+ (ours) &       & 857K  & \pmb{34.71} / \pmb{0.9297} & \pmb{30.58} / \pmb{0.8464} & \pmb{29.26} / \pmb{0.8089} & \pmb{28.78} / \pmb{0.8634} & \pmb{34.43} / \pmb{0.9490} \\
    \midrule
    VDSR \cite{kimAccurateImageSuperResolution2016}  & \multicolumn{1}{c}{\multirow{11}[2]{*}{×4}} & 666K  & 31.35 / 0.8838 & 28.01 / 0.7674 & 27.29 / 0.7251 & 25.18 / 0.7524 & 28.83 / 0.8870 \\
    CARN \cite{ahnFastAccurateLightweight2018}  &       & 1592K & 32.13 / 0.8937 & 28.60 / 0.7806 & 27.58 / 0.7349 & 26.07 / 0.7837 & 30.42 / 0.9070 \\
    IMDN \cite{huiLightweightImageSuperResolution2019}  &       & 715K  & 32.21 / 0.8948 & 28.58 / 0.7811 & 27.56 / 0.7353 & 26.04 / 0.7838 & 30.45 / 0.9075 \\
    RFDN \cite{liuResidualFeatureDistillation2020}  &       & 550K  & 32.24 / 0.8952 & 28.61 / 0.7819 & 27.57 / 0.7360 & 26.11 / 0.7858 & 30.58 / 0.9089 \\
    LatticeNet+ \cite{luoLatticeNetLightweightImage2020} &       & 777K  & 32.30 / 0.8962 & 28.68 / 0.7830 & 27.62 / 0.7367 & 26.25 / 0.7873 & - \\
    MAFFSRN \cite{muqeetMultiattentionBasedUltra2020} &       & 830K  & 32.20 / 0.8953 & 26.62 / 0.7822 & 27.59 / 0.7370 & 26.16 / 0.7887 & - \\
    SMSR \cite{wangExploringSparsityImage2021}  &       & 1006K & 32.12 / 0.8932 & 28.55 / 0.7808 & 27.55 / 0.7351 & 26.11 / 0.7868 & 30.54 / 0.9085 \\
    ESRT \cite{luTransformerSingleImage2022}  &       & 777K  & 32.19 / 0.8947 & 28.69 / 0.7833 & 27.69 / 0.7379 & 26.39 / 0.7962 & 30.75 / 0.9100 \\
    SwinIR-s \cite{liangSwinIRImageRestoration2021a} &       & 897K  & 32.44 / 0.8976 & 28.77 / 0.7858 & 27.69 / 0.7406 & 26.47 / 0.7980 & 30.92 / \underline{0.9151} \\
    HFFN (ours) &       & 867K  & \underline{32.49} / \underline{0.8988} & \underline{28.83} / \underline{0.7870} & \underline{27.74} / \underline{0.7413} & \underline{26.62} / \underline{0.8014} & \underline{31.09} / 0.9134 \\
    HFFN+ (ours) &       & 867K  & \pmb{32.56} / \pmb{0.8996} & \pmb{28.89} / \pmb{0.7881} & \pmb{27.77} / \pmb{0.7421} & \pmb{26.75} / \pmb{0.8040} & \pmb{31.34} / \pmb{0.9160} \\
    \bottomrule
    \end{tabular}%
  \label{tab:addlabel1}%
\end{table*}%
\begin{table}[t]
    \caption{Average PSNR obtained on the five benchmark datasets when the LFFB module is not used, scale factor $ \times 4$. The value of the difference from the baseline is marked in parentheses.\label{table4}}
    \centering
    \begin{tabular}{p{4.19em}cc}
    \toprule
    Method & w/o LFFB &HFFN \\
    \midrule
    Params & 831K  & 867K \\
    \midrule
    Set5  & 31.67 (-0.71dB) & 32.38 \\
    Set14 & 28.25 (-0.45dB) & 28.70 \\
    B100  & 27.33 (-0.33dB) & 27.66 \\
    Urban100 & 25.37 (-1.04dB) & 26.41 \\
    Manga109 & 29.39 (-1.42dB) & 30.81 \\
    \bottomrule
    \end{tabular}
    \end{table}
\subsubsection{Effectiveness of basic modules in low-frequency de-redundant branch}
To verify the effectiveness of each component of the low-frequency branch, we designed two sets of experiments, replacing the depth-separable convolution with a normal $3 \times 3$ convolution layer and removing the CCA module, respectively. Similarly, we used 500 epochs to train the models. From Table \ref{table2}, a comparison of the first and third rows demonstrates that the parameters can be reduced by about 30K using DSConv without affecting performance. Comparing the second row with the third row, it can be seen that CCA can improve the PSNR by about 0.1dB on the Urban100 and Manga109 datasets with only a 5K increase in the number of parameters. Therefore, in the balance between efficiency and parameter number, we used a deeply separable convolution with the CCA module.

\subsubsection{The effect of increasing the number of HFFB}
To balance the relationship between model size and performance, we set a different number of HFFBs for each LFFB module, designed models of different sizes and evaluated PSNR and SSIM on five datasets. In Table \ref{table3} it is observed that the best performance is achieved when the number of HFFBs in each LFFB is 5. The model performance does not improve when the number continues to rise. Therefore, the number of HFFBs per LFFB in our model is set to 5.

\subsubsection{Study of local feature fusion block (LFFB)}
In Table \ref{table4}, we explored the effectiveness of LFFB by removing the LFFB and designing a network directly using 30 HFFBs. The same we used 500 epochs to train our model. It can be observed that the LFFB we designed plays a vital role in the model. It allows better utilization of the extracted feature from each HFFB module.
\subsection{Comparison with State-of-the-art Methods}
In Table \ref{table6}, we provide a quantitative comparison of recent state-of-the-art lightweight SISR models with our HFFN, including VDSR\cite{kimAccurateImageSuperResolution2016}, CARN\cite{ahnFastAccurateLightweight2018}, IMDN\cite{huiLightweightImageSuperResolution2019}, RFDN\cite{liuResidualFeatureDistillation2020}, LatticeNet+\cite{luoLatticeNetLightweightImage2020}, MAFFSRN\cite{muqeetMultiattentionBasedUltra2020}, SMSR\cite{wangExploringSparsityImage2021}, ESRT\cite{luTransformerSingleImage2022}, and SwinIR-s\cite{liangSwinIRImageRestoration2021a}. The compared metrics include PSRN, SSIM, and parameter counts. As can be clearly seen, our HFFN achieves the top results on almost every dataset. In particular, our model performs much better than other models in $ \times 4$ scale image recovery, which is due to the fact that the bigger the scale factor is, the more high-frequency feature is lost in the degradation process of the image, and then our model's strategy for allocating computational resources makes more sense.
\begin{figure*}[t]
  \centering
  \subfloat{\includegraphics[width=6.5in]{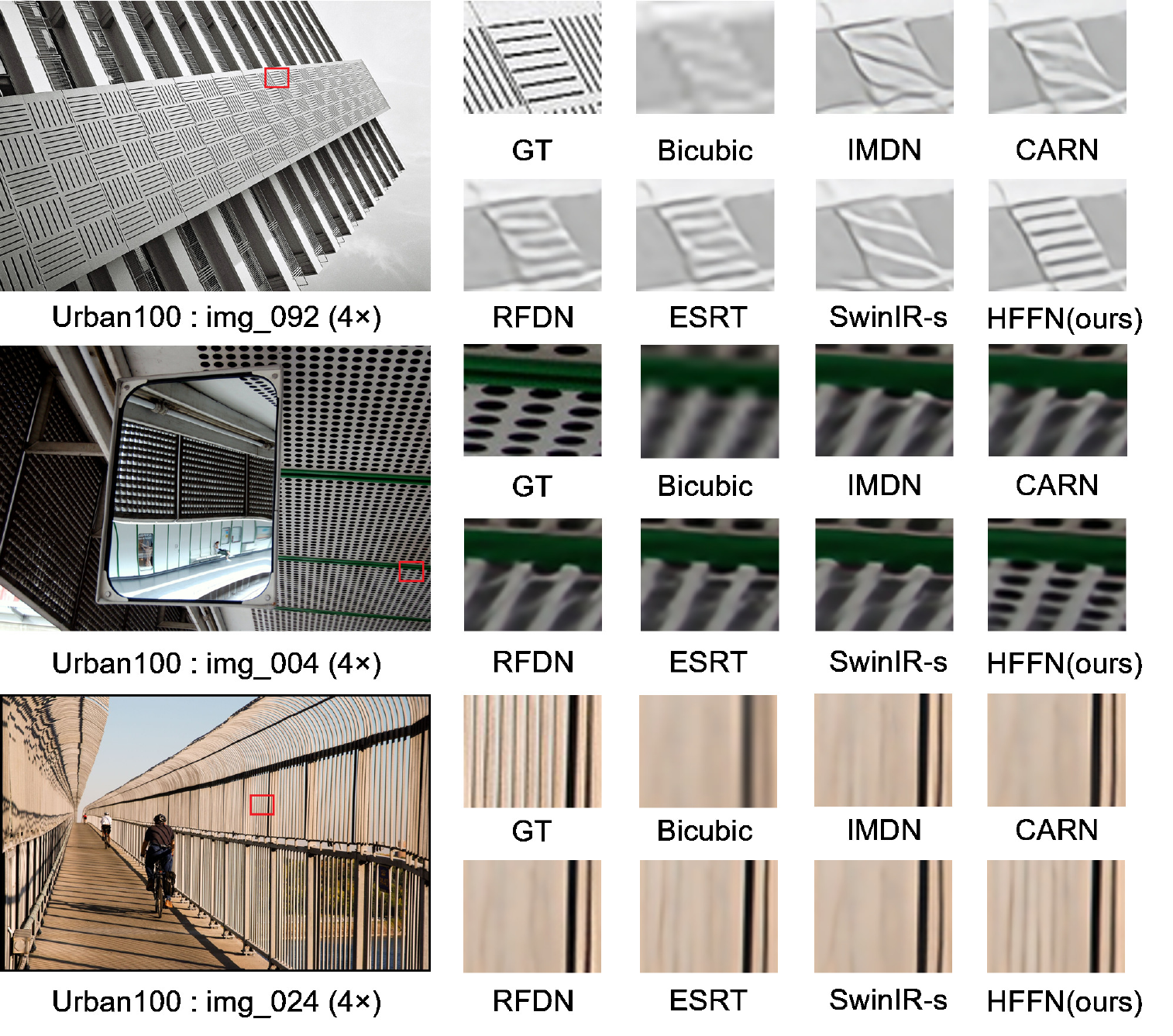}%
  \label{fig_1_case}}
  \caption{Visual comparison of HFFN with the state-of-the-art methods on ×4 SR.}
  \label{fig6}
  \end{figure*}
\begin{figure}[t]
  \centering
  \includegraphics[width=3.2in]{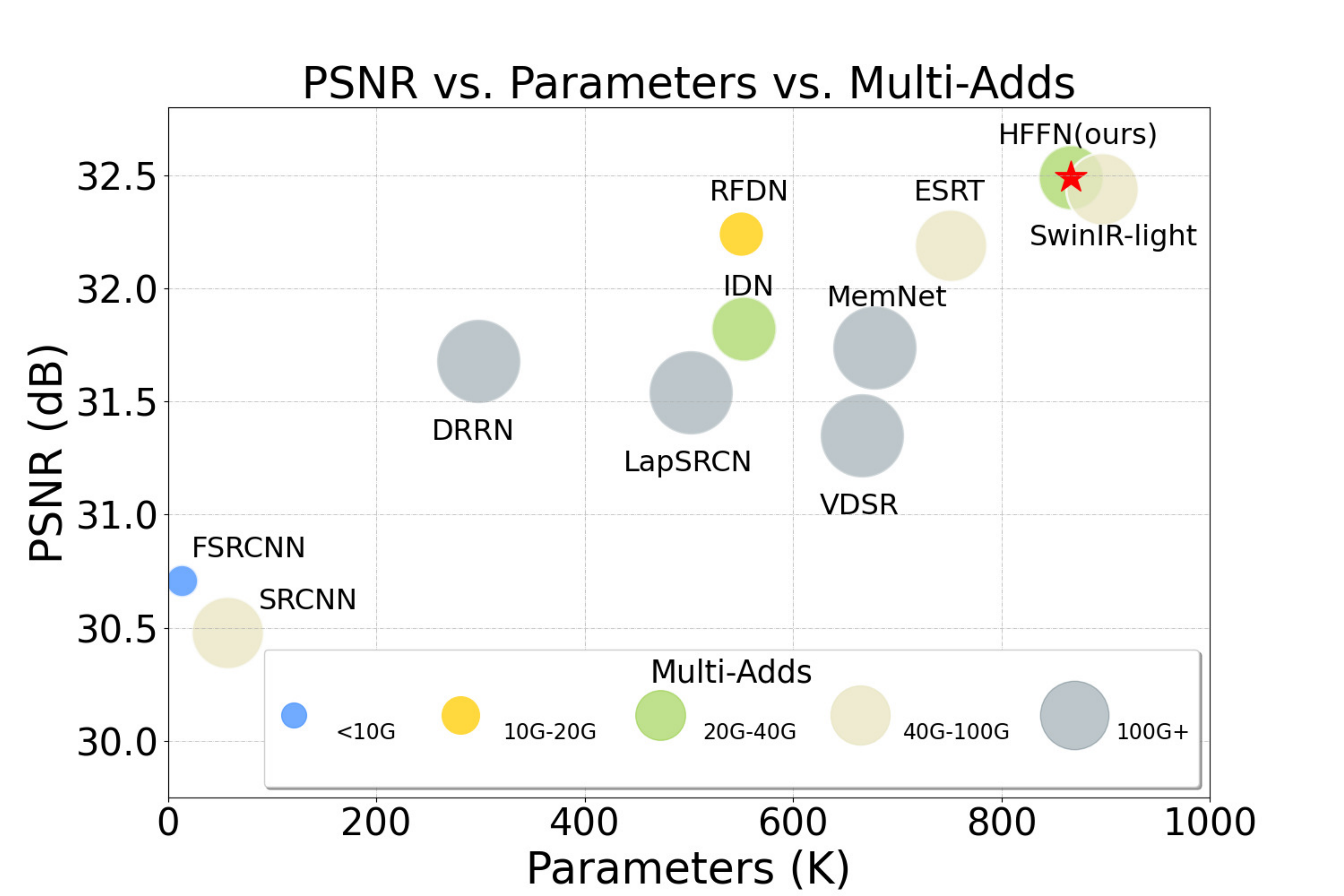}
  \caption{Performance and model complexity comparison on Set5 dataset for upscaling factor $ \times 4$.}
  \label{fig_7}
  \end{figure}

In Fig. \ref{fig6}, we provide a visual comparison of recent state-of-the-art lightweight SISR methods with HFFN, including SwinIR-s, which is considered an extremely good existing lightweight SISR model. The images indicate that the reconstructed regions of SwinIR-s are visually significantly less effective than our HFFN in regions with high-frequency features. Our HFFN is able to effectively allocate computational resources towards the recovery of these challenging high-frequency features, resulting in sharper and more visually pleasing images. These results demonstrate that our HFFN outperforms existing lightweight SISR methods (including SwinIR-s).
\subsection{Comparison of Computational Cost}
In Table \ref{table5}, we compare our HFFN with the lightweight version of the state-of-the-art method SwinIR, where we compare the number of parameters and the multiplication of our method. With the $ \times 4$ scaling factor, our HFFN can outperform SwinIR-s when the parameter count is slightly smaller, except for a slightly lower SSIM index in the Manga109 dataset. The rest of the performance is all better than SwinIR-s. Moreover, since SwinIR is designed based on the Transformer architecture and therefore far more computationally intensive than the CNN architecture, the Multi-Adds of our HFFN are only 65.7\% of the SwinIR-s. In addition, we visualize in Fig. \ref{fig_7} a comparative analysis between our parametric number and performance and the number of multiplicative operations with state-of-the-art models. It is clear that our method scores well between performance and lightness.
\begin{table}[t]
      \centering
      \caption{Comparison of computational cost.\label{table5}}
        \begin{tabular}{p{6.19em}cccc}
        \toprule
        Method & Params & Multi-Adds & Urban100 & Manga109 \\
        \midrule
          SwinIR-s & 897K  & 49.6G & 26.47 & 30.92 \\
          HFFN (ours) & 867K  & 32.6G & 26.62 & 31.09 \\
          \bottomrule
          \end{tabular}%
        \label{tab:addlabel}%
      \end{table}%
\section{Conclusion}
This paper proposes a novel lightweight High-Frequency Focus network (HFFN) for single image super-resolution (SISR). Our HFFN utilizes the Local Feature Fusion block (LFFB) as its basic module, which is composed of our designed High-Frequency Focus blocks (HFFB). The HFFN is specifically designed to allocate its limited computational resources toward recovering the high-frequency features that are more difficult to restore, thus producing sharper images. We conducted extensive experiments on five benchmark datasets, and our method outperformed the state-of-the-art lightweight SISR methods while maintaining a lower number of parameters and Multi-adds. Our results demonstrate that the HFFN is a highly effective and efficient approach for SISR.


\bibliographystyle{IEEEtran}
\bibliography{IEEEabrv,lib}

\newpage

\vfill

\end{document}